\documentclass[
showpacs,preprintnumbers,amsmath,amssymb]{revtex4}
\usepackage{amsmath}
\usepackage{graphicx}

\begin{document}

\def\be{\begin{equation}}
\def\ee{\end{equation}}
\def\bea{\begin{eqnarray}}
\def\eea{\end{eqnarray}}
\def\tr{{\rm tr}\, }
\def\nn{\nonumber \\}
\def\e{{\rm e}}

\title{Cosmological solutions in $F(T)$ gravity with the presence of spinor fields}
\author{R. Myrzakulov$^{1}$\footnote{
E-mail: rmyrzakulov@csufresno.edu} , D. S\'aez-G\'omez$^{2,3}$\footnote{
E-mail: diego.saezgomez@uct.ac.za} and P. Tsyba$^{1}$} 

\affiliation{$^{1}$Eurasian International Center for Theoretical Physics $\&$  
Department of General  and  Theoretical Physics, \\
Eurasian National University, Astana 010008, Kazakhstan\\
$^2$Astrophysics, Cosmology and Gravity Centre (ACGC) $\&$  \\ 
Department of Mathematics and Applied Mathematics, University of Cape Town, Rondebosch 7701, Cape Town, South Africa \\
$^3$Fisika Teorikoaren eta Zientziaren Historia Saila, Zientzia eta Teknologia Fakultatea,\\
Euskal Herriko Unibertsitatea, 644 Posta Kutxatila, 48080 Bilbao, Spain, EU}

\begin{abstract}
The presence of spinor fields is considered in the framework of some extensions of teleparallel gravity, where the Weitzenb\"ock connection is assumed. Some well known models as the Chaplygin gas and its generalizations are reconstructed in terms of a spinor field in the framework of teleparallel gravity. In addition, the $\Lambda$CDM model is also realized with the presence of a spinor field where a simple self-intercating term is considered and the corresponding action is reconstructed. Other cosmological solutions and the reconstruction of the gravitational action in terms of the scalar torsion is  studied.\end{abstract}
\pacs{04.50.Kd, 04.70.Bw, 04.20.Jb}

\maketitle

\section{Introduction}

Cosmology has experienced a large increase of activity over the last decades, especially due to the release of large amounts of new observational data. In this sense, the observations of Supernovae Ia by two independent groups in 1998 \cite{SN1} showed a deviation of the luminosity distance for the first time, a fact that was widely interpreted as a consequence of the accelerating expansion of the universe. From then on, other datasets  have confirmed such assumption as for instance the observations of the Cosmic Microwave Background radiation (see \cite{WMAP}) or the baryon acoustic oscillations \cite{Eisenstein:2005su}, among others. These observations along with the assumption, widely confirmed, that the universe is homogeneous and isotropic at large scales, have given rise to plenty of proposals in order to explain such anomalous behavior of the expansion. Hence, the Friedmann-Robertson-Lema\^itre-Walker (FLRW) equations, which provides a good description of the universe at large scales, should involve a fluid with negative pressure in order to produce an accelerating expansion. The behavior of this fluid, known as dark energy,  can be achieved through several ways: a positive cosmological constant,  scalar fields, vector fields or modifications of General Relativity (GR), among others (for a review on dark energy candidates, see \cite{R-DE}). Regarding the possible modifications of GR that might lead to the behavior of dark energy, the so-called $f(R)$ gravity is possibly the most well known alternative, since its simplicity by generalizing  the Hilbert-Einstein action to a more general function of the Ricci scalar, is capable to realize the late-time acceleration of the universe expansion (for a review on $f(R)$ gravity, see \cite{F(R)-gravity}). Nevertheless, other alternatives to GR have been also studied, where some other curvature invariants as for instance the Gauss-Bonnet term have been considered, leading  also to theories capable  of reproducing quite well the cosmological expansion (see \cite{MST}). \\

In addition, some extensions of the so-called teleparallel gravity have been also considered. Teleparallel gravity assumes the  Weitzenb\"{o}ck connection instead of the Levi-Civita connection to construct the gravitational action (for a review, see \cite{Teleparallelism}). In this sense, the curvature vanishes whereas the torsion scalar is not null, such that an action consisting of a linear term of the torsion leads to an equivalent theory to GR. Consequently, teleparallel gravity gives rise to the same FLRW equations than in standard GR, and dark energy is also required in order to reproduce an accelerating expansion. Nevertheless,  in analogy to $f(R)$ theories, some extensions of teleparallel gravity have been proposed under the name of $f(T)$ gravity, which are capable of reproducing the dark energy epoch (see \cite{Bengochea:2008gz,Linder:2010py,Myrzakulov:2010vz}). Hence, a considerable number of works has  analyzed this possibility, where  several cosmological models and their properties have been studied in the framework of $f(T)$ gravity (see Ref.~\cite{WY-BGLL-BGL}). Then, as in the case of $f(R)$ gravity, the aim of these theories is to reproduce the dark energy epoch with no need of any extra field, but just in terms of gravity. However, in comparison with $f(R)$ gravity, the gravitational field equations of $f(T)$ gravity are second order instead of the fourth order of $f(R)$ gravities, which simplifies the equations in comparison with $f(R)$ gravity, and leads for instance to the same gravitational waves modes as in GR \cite{Bamba:2013ooa}. Furthermore,  some other theoretical aspects have been studied within the framework of $f(T)$ gravity, as causality problems~\cite{OINC-IGO}, the realization of inflation \cite{Ferraro:2006jd}, the behavior of the cosmological perturbations~\cite{Izumi:2012qj}, some  conformal symmetries in teleparallel gravity~\cite{Bamba:2013jqa} or the local Lorentz invariance of the theory~\cite{Li:2010cg}. \\

On the other hand, the possibility of reproducing dark energy by means of spinor fields has been also considered in the literature \cite{Ribas:2005vr}. In this sense, any cosmological solution can be reconstructed by the appropriate self-interacting term of the spinor field in the framework of GR and even within higher order theories of gravity as $f(R)$ gravity~\cite{Fabbri:2010pk}, so that the cosmological history might be explained by sources of Dirac fields. Furthermore, some inflationary models with the presence of spinor fields have been proposed~\cite{ArmendarizPicon:2003qk}, as well as models that mimic some well known dark energy models as for instance, quintessence or the Chaplygin gas model~\cite{Krechet:2010zza}. In addition, some cosmological solutions have been analyzed with the presence of spinor fields in  gravitational theories with non-null torsion~\cite{Carloni:2013hna}. \\

In this paper, we deal with the presence of spinor fields in the framework of $f(T)$ gravity, where the Weitzenb\"{o}ck connection is assumed. Several dark energy models are reconstructed in teleparallel gravity with the presence of an additional trivial term in the action. In this sense, the self-interacting term of the spinor field is reconstructed in order to reproduce the behavior of the Chaplygin gas, as well as some of its generalizations. Then, by assuming a usual  self-interacting term of the spinor field, the corresponding $f(T)$ gravity is obtained, where it is shown that $\Lambda$CDM can be well realized. Some other cosmological solutions are also explored.\\

The paper is organized as follows: in Section \ref{fTeqs}, the main aspects of $f(T)$ gravity are briefly reviewed. Section \ref{Tspinor} deals with the presence of Dirac fields in the framework of teleparallel gravity, where the Chaplygin gas models are reconstructed in terms of a spinor field. Section \ref{fTspinor} is aimed to the reconstruction of the corresponding $f(T)$ action for some particular cosmological solutions as $\Lambda$CDM model. Finally, Section \ref{conclusions}  summarizes the results of the paper.

\section{$F(T)$  gravity}
\label{fTeqs}

Let us start by reviewing the main aspects of $F(T)$ gravities, which refers in this case to an extension of the well known Teleparallel gravity. Teleparallel theories are described by the vierbein vectors  $e^{A}$, defined in the tangent space of a particular manifold by the components   $e^{i}_{\mu}$. Moreover, the Weitzenb\"{o}ck connection is assumed instead of the Levi-Civita connection,
\begin{eqnarray}
\tilde{\Gamma}^{\alpha}_{\mu\nu}=e_{i}^{\;\;\alpha}\partial_{\nu}e^{i}_{\;\;\mu}=-e^{i}_{\;\;\mu}\partial_{\nu}e_{i}^{\;\;\alpha}\label{co}\; ,
\label{WC}
\end{eqnarray}
which leads to a null covariant derivative of the vierbein, $\nabla_{\mu}e^{i}_{\;\;\nu}=0$, in analogy to the pure curvature gravity, where the Levi-Civita connection leads to a null divergence of the metric tensor.  The action for $F(T)$  gravity is given by  (see, e.g. \cite{Bengochea:2008gz,Linder:2010py,Myrzakulov:2010vz})
\begin {equation}
S=\int d^{4}xe[\frac{1}{2\kappa^{2}}F(T)+L_m ]\ ,
\end{equation} 
where $T$ is the torsion scalar, $e=\det{(e^{i}_{\mu})}=\sqrt{-g}$ and $L_m$ stands for the matter Lagrangian. The components $e^{i}_{\mu}$ are defined by the vierbein vector field $\textbf{e}_{A}$ in a coordinate basis, that accomplishes $\textbf{e}_{A}\equiv e^{\mu}_{A}\partial_{\mu}$. Note that in teleparallel gravity, the dynamical variable is the vierbein field $\textbf{e}_{A}(x^{\mu})$. Then,  the variation of the action with respect to the vierbein  leads to  the following field equations
\begin {equation}
[e^{-1}\partial_{\mu}(eS^{\mu\nu}_{i})-e^{\lambda}_{i}T^{\rho}_{\mu\lambda}S^{\nu\mu}_{\rho}]F_{T}+S^{\mu\nu}_{i}(\partial_{\mu}T)F_{TT}+\frac{1}{4}e^{\nu}_{i}F=\frac{1}{2}\kappa^2e^{\rho}_{i}T^{\nu}_{\rho}.
\end{equation}
And the torsion scalar  
$T$  is given by
\begin{equation}
T=S^{\mu\nu}_{\rho}T^{\rho}_{\mu\nu}
\label{te}
\end{equation}
with
\begin{equation}
	S_\rho\,^{\mu\nu}=\frac{1}{2}(K^{\mu\nu}\,_\rho+\delta^\mu_\rho T^{\theta\nu}\,_\theta-\delta^\nu_\rho T^{\theta\mu}\,_\theta).
\end{equation}
Whereas the contorsion tensor is defined as
\begin{equation}
K^{\mu\nu}\,_\rho=-\frac{1}{2}(T^{\mu\nu}\,_\rho-T^{\nu\mu}\,_\rho-T_\rho\,^{\mu\nu})
\label{contor}
\end{equation}
and the torsion tensor is defined as the antisymetric part of the connection (\ref{WC}), 
\begin{equation}
T^\lambda_{\mu\nu}=\stackrel{w}{\Gamma}^\lambda_{\nu\mu}-\stackrel{w}{\Gamma}^\lambda_{\mu\nu}=e^\lambda_i(\partial_\mu e^i_\nu-\partial_\nu e^i_\mu).
\label{tor}
\end{equation}
Then, the vierbein vector fields are related with the metric through
\begin{equation}
g_{\mu\nu}(x)=\eta_{ij}e^i_\mu(x)e^j_\nu(x), 
\label{metric-vierbein}
\end{equation}
where $ {\bf e}_i \cdot {\bf e}_j=\eta_{ij}$ and $\eta_{ij}=diag(1, -1, -1, -1)$. Here, a flat homogeneous and isotropic FLRW universe is considered, whose metric is given by
\begin{equation}
ds^{2}=dt^{2}-a(t)^{2}\sum^{3}_{i=1}(dx^{i})^{2},
\label{metricFLRW}
\end{equation}
where $t$ is  cosmic time. Along the present work, a diagonal set of tetrads is assumed, which is expressed as follows
\be
e^{i}_{\;\;\mu}=\text{diag}\left(1,a,a,a\right)\ , \quad e_{i}^{\;\;\mu}=\text{diag}\left(1,\frac{1}{a},\frac{1}{a},\frac{1}{a}\right)\ .
\label{FLRWtetrad}
\ee
The set of tetrads (\ref{FLRWtetrad}) corresponds to the metric (\ref{metricFLRW}) trough the relation (\ref{metric-vierbein}), whereas $e=a^3$ is the determinant of the matrix. The non-null components of the torsion tensor (\ref{tor}) and the contorsion tensor (\ref{contor}) for this choice of the vierbein are
\begin{eqnarray}
T^{1}_{\;\;01}=T^{2}_{\;\;02}=\,T^{3}_{\;\;03}=K^{01}_{\;\;\;\;1}=K^{02}_{\;\;\;\;2}=K^{03}_{\;\;\;\;3}=H(t)\; ,\label{torsiontype3}
\end{eqnarray}
whereas the components of $S_{\alpha}^{\;\;\mu\nu}$ in (\ref{s}) are given by
\begin{eqnarray}
S_{1}^{\;\;10}=S_{2}^{\;\;20}=S_{3}^{\;\;30}=H(t)\,\;.\label{tensortype3}
\end{eqnarray}
Finally the torsion scalar (\ref{te}) yields
\begin{eqnarray}
T=-6H^2(t)\; \label{torsionScalar1}.
\end{eqnarray}
Then, we can write the modified Friedmann equations and the continuity equation as follows
\begin{equation}
	-2TF_{T}+F=2\kappa^2 \rho_m\ , \quad -8\dot{H}TF_{TT}+(2T-4\dot{H})F_{T}-F=2\kappa^2p_m, 
\label{fTFLRW}
\end{equation}
\begin{equation}
	\dot{\rho}_m+3H(\rho_m+p_m)=0\ ,
	\label{contieq}
\end{equation}
which can be rewritten in a more appropriate form as
\begin{equation}
	-T-2Tf_{T}+f=2\kappa^2 \rho_m\ , \quad 	-8\dot{H}Tf_{TT}+(2T-4\dot{H})(1+f_{T})-T-f=2\kappa^2p_m, 
	\label{fTFLRW1}
\end{equation}
where we have assumed the action
\begin {equation}
S=\int d^{4}xe[\frac{1}{2\kappa^{2}}(T+f(T))+L_{m}],
\label{action1}
\end{equation} 
and $f=F-T$, so that the extra components enclosed in $f(T)$ can be seen as corrections to the usual teleparallel action.  Moreover, one might define an effective density and pressure for the extra terms of the FLRW equations (\ref{fTFLRW1}) as
\begin{eqnarray}
	\rho_T&=&\kappa^{-2}(Tf^{'}-0.5f),\nn
p_T&=&\kappa^{-2}[4\dot{H}Tf^{"}+(2\dot{H}-T)f^{'}+0.5f]=-\rho_T+2\kappa^{-2}\dot{H}[2Tf^{"}+f^{'}]\ ,
\label{densities}
	\end{eqnarray}
Then, the equations (\ref{fTFLRW1}) become the usual FLRW equations of teleparallel gravity, or equivalently General Relativity,
  	\begin{eqnarray}
	3H^2-\kappa^2\rho=0\ ,\quad 2\dot{H}+3H^2+\kappa^2p=0\ ,
	\end{eqnarray}
	where
	\begin{equation}
	\rho=\rho_m+\rho_T,\quad p=p_m+p_T 
\end{equation}
Here $\rho_m$ and $p_m$ are the energy and pressure of the matter content respectively. By defining $F=2E$ and $\kappa^2=1$, the equations (\ref{fTFLRW1}) yield
	\begin{eqnarray}
	-2TE_{T}+E- \rho_m=0\ ,\quad	-8\dot{H}TE_{TT}+(2T-4\dot{H})E_{T}-E-p_m=0\ , 
\end{eqnarray}
which have the following general solution
	\begin{equation}
	E=C\sqrt{T}-\frac{1}{2}\sqrt{T}\int\frac{\rho_m}{T^{3/2}}dT 
	\label{E1}
\end{equation}
where $C$ is an integration constant. 
Hence, by expressing $\rho_m$ in terms of the torsion, the corresponding $f(T)$ action can be reconstructed for a particular cosmological evolution.

\section{Teleparallel  gravity with the presence of spinor fields}
\label{Tspinor}

Let us start by studying the particular case,
  \be 
  f\left(T\right)=c\sqrt{T}\ ,
  \label{action2}
  \ee
where $c=const$, and the action (\ref{action1}) turns out  the teleparallel action plus the term (\ref{action2}) $F(T)=T+c\sqrt{T}$, whose second term  is cancelled in the lhs of the FLRW equations.  In this case, as follows from (\ref{densities}), the effective densities are null $\rho_T=p_T=0$. By  assuming  the matter Lagrangian of a complex spinor field,  

\begin {equation}
L_m = Y-V(\bar{\psi},\psi)\ ,
\label{DiracL}
\end{equation} 
where $\psi=(\psi_1, \psi_2, \psi_3, \psi_4)^{T}$  is a spinor function  and $\bar{\psi}=\psi^{\dagger}\gamma^0$ is its adjoint function, with the dagger representing complex conjugation. The canonical kinetic term for the spinor field is defined as
\begin {equation}
Y=\frac{1}{2}i[\bar{\psi}\Gamma^{\mu}D_{\mu}\psi-(D_{\mu}\bar{\psi})\Gamma^{\mu}\psi]\ ,
\label{kinetic1}
\end{equation}
where   $ D_{\mu}$ is a covariant derivative. Note that the spinor fields are treated here as classical commuting fields.  In general, the field equations corresponding to the action (\ref{action2}) and the Lagrangian (\ref{DiracL}) have a very complicated form, so here we  focus on a flat  FLRW metric with the presence of the spinor field described by the Lagrangian (\ref{DiracL}). Recall that the set of vierbeins  is chosen to be diagonal,
	\begin{equation}
	(e_a^\mu)=\text{diag}(1,1/a,1/a,1/a),\quad 
(e^a_\mu)=\text{diag}(1,a,a,a)\ ,
\end{equation}
which corresponds to the metric (\ref{metricFLRW}) as can be shown by (\ref{metric-vierbein}). The preliminary set-up for writing the equations of motion is now complete, so for  the FLRW metric (\ref{metricFLRW}), the equations corresponding to the action (\ref{action2}) and the spinor Lagrangian (\ref{DiracL}) yield \cite{ArmendarizPicon:2003qk}
	\begin{eqnarray}
	3H^2-\rho&=&0,  \quad 2\dot{H}+3H^2+p=0,\nn
				\dot{\psi}+\frac{3}{2}H\psi+i\gamma^0V_{\bar{\psi}}&=&0,\quad \dot{\bar{\psi}}+\frac{3}{2}H\bar{\psi}-iV_{\psi}\gamma^{0}=0,\nn
	\dot{\rho}+3H(\rho+p)&=&0,
	\label{TeleparallelDIrac}
	\end{eqnarray} 
where  the dot denotes derivatives with respect to the cosmic time $t$. Here the kinetic term, the energy density  and  the pressure  take the form
\begin{equation}
 Y=\frac{1}{2}i(\bar{\psi}\gamma^{0}\dot{\psi}-\dot{\bar{\psi}}\gamma^{0}\psi)
  \end{equation}
  and
  \begin{equation}
  \rho=\rho_m=V,\quad
p=p_m=Y-V,
  \end{equation}
 respectively. Then, for a particular spinor potential, the Hubble parameter can be obtained. In the following, we study some cases that yield to late-time acceleration, and where the spinor field behaves as some well known dark energy models.
 
 \subsection{The spinor Chaplygin gas model} 
 
 As a first example, let us consider  the original Chaplygin gas, whose equation of state is defined as
   \begin{equation}
 p=-\frac{A}{\rho},
  \end{equation}
where $A$ is a positive constant. The cosmological model based on the Chaplygin gas
was proposed for the first time in \cite{Kamenshchik:2001cp} as an alternative to the quintessences models for dark energy. Let us assume a potential for the spinor field which has the special form $V=V(\bar{\psi}, \psi)=V(u)$, where $u=\bar{\psi} \psi$ is a scalar, and no other matter contribution.  Then, in terms of the scale factor $a$, the system (\ref{TeleparallelDIrac}) has the following solution
  \begin{eqnarray}
	H&=&\pm 3^{-\frac{1}{2}}(A+Ba^{-6})^{\frac{1}{4}},\nn
		\rho&=&(A+Ba^{-6})^{\frac{1}{2}},\nn
				p&=&-A(A+Ba^{-6})^{-\frac{1}{2}},\nn
				\psi_j&=&c_ja^{-\frac{3}{2}}e^{-iD},\quad j=1,2,\nn
				\psi_l&=&c_la^{-\frac{3}{2}}e^{iD},\quad l=3,4,\nn
					V&=&(A+Ba^{-6})^{\frac{1}{2}}.	
	\end{eqnarray} 
Here $B, c_j, c_l, c$ are constants, being $c=|c_1|^2+|c_2|^2-|c_3|^2-|c_4|^2$ and
	 \begin{equation}
D=\mp 0.5\sqrt{3}Bc^{-1}a^{-2}(A+Ba^{-6})^{-\frac{3}{4}}.
  \end{equation}
Whereas  the scale factor is given by \cite{Kamenshchik:2001cp}
	 \begin{equation}
t=\frac{1}{6\sqrt[4]{A}}\left(\ln\frac{\sqrt[4]{A+Ba^{-6}}+\sqrt[4]{A}}{\sqrt[4]{A+Ba^{-6}}-\sqrt[4]{A}}-2\arctan\sqrt[4]{1+A^{-1}Ba^{-6}}\right).
  \end{equation}
And the EoS parameter yields
   \begin{equation}
\omega=-1+\frac{B}{B+Aa^{6}}.
  \end{equation}
  Finally the  expressions for the potential, and the kinetic term can be expressed in terms of $u$:
  \begin{eqnarray}
V&=&\sqrt{A+Bc^{-2}u^2}, \nn
 Y&=&Ba^{-6}(A+Ba^{-6})^{-\frac{1}{2}},\nn
 u&=&ca^{-3},\nn
 \end{eqnarray}
 Hence, the  Chaplygin gas is reproduced in terms of a Dirac field for the action $F(T)=T+c\sqrt{T}$, a possibility explored previously in the Einstein-Dirac theory~\cite{Saha:2009zm}. 
  
  \subsection{The  generalized  Chaplygin gas model}

The previous analysis can be also extended to the so-called generalized  Chaplygin gas, whose EoS has the following form \cite{Kamenshchik:2001cp} 

   \begin{equation}
 p=-\frac{A}{\rho^\alpha},
  \end{equation}
where $A$ is a positive constant. For simplicity, we assume again a spinor  potential  given by $V=V(\bar{\psi}, \psi)=V(u)$. Then,  the system of equations (\ref{TeleparallelDIrac}) has the following solution
  \begin{eqnarray}
	H&=&\pm 3^{-\frac{1}{2}}[A+Ba^{-3(\alpha+1)}]^{\frac{1}{2(\alpha+1)}},\nn
		\rho&=&[A+Ba^{-3(\alpha+1)}]^{\frac{1}{\alpha+1}},\nn
				p&=&-A[A+Ba^{-3(\alpha+1)}]^{-\frac{\alpha}{\alpha+1}},\nn
				\psi_j&=&c_ja^{-\frac{3}{2}}e^{-iD},\quad j=1,2,\nn
				\psi_l&=&c_la^{-\frac{3}{2}}e^{iD},\quad l=3,4,\nn
					V&=&[A+Ba^{-3(\alpha+1)}]^{\frac{1}{\alpha+1}}\ .	
	\end{eqnarray}	
Here $B, c_j, c_l, c$ are constants, and $c=|c_1|^2+|c_2|^2-|c_3|^2-|c_4|^2$, while $D$ is expressed as
\begin{equation}
D=\mp \sqrt{3}Bc^{-1}\int a^{-3(1+\alpha)+3}\left[A+Ba^{-3(1+\alpha)}\right]^{-\frac{0.5+\alpha}{1+\alpha}}da.
  \end{equation}
In this case, the EoS parameter can be written in terms of the scale factor as
   \begin{equation}
\omega=-1+\frac{B}{B+Aa^{3(\alpha+1)}}.
  \end{equation}
 Then, as in the above case, the expressions for the potential, the kinetic term and $u$ are given by
 \begin{eqnarray}
V&=&\left[A+Bc^{-(1+\alpha)}u^{1+\alpha}\right]^{\frac{1}{1+\alpha}}, \nn
Y&=&Ba^{-3(1+\alpha)}\left[A+Ba^{-3(1+\alpha)}\right]^{-\frac{\alpha}{1+\alpha}},\nn
u&=&ca^{-3}\nn
 \end{eqnarray}
 Hence, the generalized Chaplygin gas can be also reproduced by spinors, where the particular potential for this case is reconstructed.
 
  \subsection{The  modified Chaplygin gas model}
 
Finally, let us consider the modified Chaplygin gas model for  dark energy model. In this case, the EoS is given by a generalization of the previous ones, 
   \cite{Benaoum:2002zs}
    \begin{equation}
p=E\rho-\frac{A}{\rho^\alpha},
\label{EoS3}
  \end{equation}
 where $A$ and $E$ are positive constants and $0\leq\alpha\leq 1.$  Then, by the continuity equation $\dot{\rho}+3H(\rho+p)=0$, the energy density for this particular EoS (\ref{EoS3}) evolves as \cite{Benaoum:2002zs}
\begin{equation}
\rho=\left[A(1+E)^{-1}+Ba^{-3(1+\alpha)(1+E)}\right]^{\frac{1}{1+\alpha}},
  \end{equation}
 where $B$ is an integration constant.   As above, we assume again a potential described by $V=V(\bar{\psi}, \psi)=V(u)$.   The corresponding solution of the set of equations (\ref{TeleparallelDIrac})  in terms of the scale factor $a$ yields,
   \begin{eqnarray}
	H&=&\pm 3^{-\frac{1}{2}}\left[A(1+E)^{-1}+Ba^{-3(1+\alpha)(1+E)}\right]^{\frac{1}{2(1+\alpha)}},\nn
		\rho&=&\left[A(1+E)^{-1}+Ba^{-3(1+\alpha)(1+E)}\right]^{\frac{1}{1+\alpha}},\nn
				p&=&[EBa^{-3(1+\alpha)(1+E)}-A(1+E)^{-1}]\left[A(1+E)^{-1}+Ba^{-3(1+\alpha)(1+E)}\right]^{-\frac{\alpha}{1+\alpha}},\nn
				\psi_j&=&c_ja^{-\frac{3}{2}}e^{-iD},\quad j=1,2,\nn
				\psi_l&=&c_la^{-\frac{3}{2}}e^{iD},\quad l=3,4,\nn
					V&=&\left[A(1+E)^{-1}+Ba^{-3(1+\alpha)(1+E)}\right]^{\frac{1}{1+\alpha}}\ ,	
	\end{eqnarray} 
where
	 \begin{equation}
D=\mp \sqrt{3}B(1+E)c^{-1}\int a^{-3(1+\alpha)(1+E)+3}\left[A(1+E)^{-1}+Ba^{-3(1+\alpha)(1+E)}\right]^{-\frac{0.5+\alpha}{1+\alpha}}da.
  \end{equation}
	
And the EoS parameter evolves as
   \begin{equation}
\omega=\frac{EBa^{-3(1+\alpha)(1+E)}-A(1+E)^{-1}}{A(1+E)^{-1}+Ba^{-3(1+\alpha)(1+E)}}\ .
  \end{equation}
 Whereas the corresponding  expressions for the potential, the kinetic term and $u$ read as:
 \begin{eqnarray}
V&=&\left[A(1+E)^{-1}+Bc^{-(1+\alpha)(1+E)}u^{(1+\alpha)(1+E)}\right]^{\frac{1}{1+\alpha}}, \nn
Y&=&(1+E)Ba^{-3(1+\alpha)(1+E)}\left[A(1+E)^{-1}+Ba^{-3(1+\alpha)(1+E)}\right]^{-\frac{\alpha}{1+\alpha}},\nn
u&=&ca^{-3},\nn
 \end{eqnarray} 
  
Then, by using the reconstruction techniques shown in the previous section,  the spinor potential can be reconstructed for any cosmological model. Here, we have illustrated such procedure by reconstructing the spinor potential for several kinds of the Chaplygin gas model, leading to an alternative to the usual dark energy candidates.


\section{$F(T)$  gravity and spinors fields: reconstructing cosmological solutions}
\label{fTspinor}

Let us now consider a more general action $F(T)$ with the presence of a spinor field as well as other matter content. In such a case, the FLRW equations are given by (\ref{fTFLRW})
\begin{equation}
	-2TF_{T}+F=2\kappa^2\sum_{i} \rho_i\ , \quad -8\dot{H}TF_{TT}+(2T-4\dot{H})F_{T}-F=2\kappa^2\sum_i p_i, 
	\label{ft0}
\end{equation}
where the sum in the matter sector of the equations is given by the contributions of all species present in the universe. By assuming a pressureless fluid and a spinor field described by the Lagrangian (\ref{DiracL}), the energy and pressure densities yield
\be
\rho_{Total}=\rho_m+\rho_{\psi}\ , \quad p_{Total}=p_\psi\ .
\label{ft1}
\ee
where recall that $\rho_{\psi}=V$ and $p_{\psi}=Y-V$, being $Y$ the kinetic term of the spinor field (\ref{kinetic1}) and $V$ the spinor potential, a function to be determined. Hence, note that by assuming an ansatz solution $H(t)$ and a particular spinor potential $V$, the corresponding action $F(T)$ that reproduces such solution can be reconstructed. In order to illustrate this procedure, let us consider the $\Lambda$CDM evolution as the solution for the Hubble parameter,
\be
H=\frac{\kappa}{3}\rho_{0} a^3+\frac{\Lambda}{3}=\frac{\kappa^3}{3}\rho_{0} (1+z)^3+\frac{\Lambda}{3}\ .
\label{LCDM1a}
\ee
Here, $\rho_0$ and $\Lambda$ are constants to be matched with the observations. For simplicity, the following spinor potential is considered,
\be
V(\bar{\psi},\psi)=m \bar{\psi}\psi\ .
\label{ft2}
\ee
Then, the Dirac equation yields,
\be
a H(a) \psi'(a)+\frac{3}{2}H(a)\psi(a)+i m\gamma^0\psi(a)=0\ .
\label{ft3}
\ee
It is straightforward to solve this equation for the Hubble parameter (\ref{LCDM1a}), whose solution is given by
\bea
\psi_j&=&c_j a^{-3/2}\left(\frac{\kappa^2}{3}\rho_0+\frac{\Lambda}{3}a^3\right)^{-i\frac{m}{\Lambda}}\ , \quad j=1,2\ , \nn
\psi_k&=&c_k a^{-3/2}\left(\frac{\kappa^2}{3}\rho_0+\frac{\Lambda}{3}a^3\right)^{i\frac{m}{\Lambda}}\ , \quad k=3,4\ . 
\label{ft4}
\eea
And the potential (\ref{ft2}) in terms of the scale factor can be rewritten as
\be
V=\frac{m\mathcal{C}^2}{a^3}\ ,
\label{ft4}
\ee
where $\mathcal{C}^2=c_1^2+c_2^2-c_3^2-c_4^2$. Then, the first FLRW equation in (\ref{ft0}) becomes
\be
-2TF_{T}+F=2\kappa^2 \left(V+\rho_{m0}a^{-3}\right)=2a^{-3}\kappa^2 \left(m\mathcal{C}^2+\rho_{m0}\right)\ ,
\label{ft5}
\ee
 where we have used the continuity equation for the pressureless fluid. Then, by using $T=-6H^2$, the equation (\ref{ft5}) can be written in terms of the torsion scalar,
 \be
-2TF_{T}+F+\frac{m\mathcal{C}^2+\rho_{m0}}{\rho_0}\left(T+2 \Lambda\right)\ ,
\label{ft6}
\ee 
whose solution leads to the action
\be
F(T)=\frac{m\mathcal{C}^2+\rho_{m0}}{\rho_0}(T-2\Lambda)+k_1\sqrt{-T}\ .
\label{ft7}
\ee
This action reduces to the action of Teleparallel gravity with a cosmological constant and the trivial term $\sqrt{-T}$. Nevertheless, note that for a more complex spinor potential, the action $F(T)$ would contain more complex functions of the torsion scalar $T$. Nevertheless, in general an explicit expression for the action $F(T)$ will not be possible to reconstruct and numerical resources will be required.\\

Let us consider another class of important cosmological solutions, given by powers of the scale factor,
\be
H\propto a^n\ .
\label{ft8}
\ee
By assuming the potential (\ref{ft2}), the solution of the Dirac equation (\ref{ft3}) yields
\bea
\psi_j&=&c_j a^{-3/2}\exp\left(i \frac{m}{n}a^{-n}\right)\ , \quad j=1,2\ , \nn
\psi_k&=&c_k a^{-3/2}\exp\left(-i \frac{m}{n}a^{-n}\right)\ , \quad k=3,4\ . 
\label{ft9}
\eea
Whereas the spinor potential is given by $V=\frac{m\mathcal{C}^2}{a^3}$ in terms of the scale factor. Then, the first FLRW equation (\ref{ft0}) turns out
\be
 -2TF_{T}+F=2\kappa^2 \frac{m\mathcal{C}}{a^3}=2\kappa^2m\mathcal{C} \left(-\frac{T}{6}\right)^{-\frac{3}{2 n}}\ ,
 \label{ft10}
 \ee
 Note that for this example, we have only considered the contribution of the spinor fields, neglecting any other matter contribution. Finally, the action that satisfies the FLRW (\ref{ft10}) is given by
 \be
 F(T)=2\kappa^2 \mathcal{C} m \left(-\frac{T}{6}\right)^{-\frac{3}{2n}}+k_1\sqrt{-T}\ .
 \label{ft11}
 \ee
 Here $k_1$ is an integration constant. Hence, any power law solution (\ref{ft8}) can be realized by spinors with the potential (\ref{ft2}) when assuming the appropriate action (\ref{ft11}), so spinors may reproduce, together with the extra terms in the action, a  matter-like evolution $(n=-3/2)$,  radiation-like cosmology $(n=-2)$ or even an accelerating expansion $(n>-1)$.

\section{Conclusions}
\label{conclusions}

In this work, the presence of Dirac fields in the framework of $f(T)$ gravity has been analyzed. Specifically, the reconstruction of some particular solutions has been performed. It has been shown that for FLRW metrics, the presence of terms as $\sqrt{-T}$ in the action reduces to the usual equations of teleparallel gravity, or equivalently to the FLRW equations in GR. In this context, the self-interacting term of the spinor field has been reconstructed following some reconstruction techniques in order to describe some well known models of dark energy. In this sense, several kinds of the Chaplygin gas model has been studied. It is well known that the Chaplygin gas models can lead to a good description of the universe evolution. Then, in this paper we have shown that every of the Chaplygin gas models can be reproduced by the appropriate   spinor potential, leading to an alternative description of the mentioned models, which are usually expressed in terms of scalar fields instead of other fields.  \\

In addition, a simple spinor model has been considered with an arbitrary $f(T)$ action. The $\Lambda$CDM solution for the Hubble parameter has been been analyzed, where has been shown that the spinor energy density would behave as a pressureless fluid, whereas the $f(T)$ action leads to the one of teleparallel gravity with the presence of a cosmological constant. Nevertheless, this is not the case for other kind of solutions, as power law expansions, where the corresponding $f(T)$ action contains additional terms in the action. Furthermore, note that a simple spinor potential has been considered in this case, such that any generalization to other more complex potentials would give rise to more complex  $f(T)$ actions. \\

Hence, we have shown a reconstruction procedure in the framework of extended teleparallel gravities with the presence of spinor fields, which can lead to reproduce realistic cosmological scenarios in the context of $f(T)$ gravities.

\section*{Acknowledgments}
 
D. S.-G. acknowledges the support from the University of the Basque Country, Project Consolider CPAN Bo. CSD2007-00042 and the URC financial support from the University of Cape Town (South Africa).

\end{document}